# The Amplification in Strong Electric and Magnetic Fields

I.V. Dovgan*

Department of Physics, Moscow State Pedagogical University, Moscow 119992, Russia.

We consider the amplification in strong electric (magnetic) fields that are uniform along the direction of the electron motion and are not uniform in the transverse direction. It is shown that in such a system the gain is increased compared with that of a magnetic undulator by a value of the order of $\gamma$ ($\gamma = \varepsilon / m_e c^2$ is the relativistic factor). Numerical estimates are presented that determine the possibility of amplification of a test wave in the optical and near infrared bands.

## 1. INTRODUCTION

The motion and radiation of relativistic electrons in stationary spatially periodic magnetic fields (undulators) has been investigated in detail in connection with the problem of free-electron lasers (FEL) [1-2]. An expressions were obtained for the amplification coefficient of an electromagnetic wave propagates along and at an angle to the direction of motion of the electron beam. It turns out that in all realistic cases the gain for electrons of energy $\varepsilon > 100 MeV$ is so small in the optical band, that the probability of generating the corresponding coherent radiation in undulators becomes quite problematic. It seems vital therefore to seek for other ways of generating coherent radiation in the soft ultraviolet in the FEL regime, without the use of undulators. One such method may be the use of ultrarelativistic electrons in the fields, that are uniform along the direction of motion but are not uniform in a direction transverse to the stationary electric and magnetic fields. The stimulated absorption and emission probabilities are found in such system [3,4]. Fedorov, Oganesyan and Prokhorov [5-7] have presented transverse channeling of electrons in intense standing light wave. Many aspects of amplification in FEL are presented in [8-59].

We calculate in the present paper in the ultrarelativistic limit the gain for a test wave propagating both along the electron beam and at an angle to it. An expression is obtained for the gain as a function of the length of the interaction region, of the angle of entry of the particles into the field, of the field intensity, of the polarization of the test wave, and others. We note that fields having a configuration close to those considered in the present paper are produced in electric and magnetic quadrupole lenses.

---

*dovganirv@gmail.com

## 2. BASIC EQUATIONS

The rate of amplification of the wave is determined by the difference between the induced-emission and absorption probabilities. Using expressions (24) of [3] and integrating with respect with the aid of (25) of [3] we obtain for the difference between the total probabilities of emitting and absorbing per unit time a quantum of frequency $\omega_1$

$$\Delta w_e^{(s=1)} = \frac{\pi}{4}\left(\frac{eA_{02}}{z_0}\right)^2 |\mathbf{e}_2\mathbf{e}_z|^2 \frac{\gamma^2}{\varepsilon\omega_1}\left[-f(\varepsilon)+\omega_1\frac{df}{d\varepsilon}\sum_n n|a_n|^2\right]. \tag{1}$$

In the derivation of (1) for the difference between the values of the function $f(\varepsilon)$ at the points $\varepsilon_e$ and $\varepsilon_a$ we have used the approximate equation

$$f(\varepsilon_e)-f(\varepsilon_a) \approx 2\Delta\varepsilon\frac{df}{d\varepsilon} \approx \frac{\omega_s}{(\delta\varepsilon)^2}, \quad \delta\varepsilon \gg \omega_s, \tag{2}$$

in which the derivative is calculated at the point $\varepsilon = \varepsilon_0$.

In accordance with the definition (12) of [3], the level populations at $n \gg 1$ are

$$|a_n|^2 \approx \frac{1}{\pi\eta_0}\frac{1}{(2\pi)^{1/2}}\left[\frac{\sin(\beta+(2n)^{1/2})\eta_0}{\beta+(2n)^{1/2}}+(-1)^n\frac{\sin(\beta-(2n)^{1/2})\eta_0}{\beta-(2n)^{1/2}}\right]$$

$$\beta \equiv p_\perp z_0 = \pm(2\varepsilon_\perp/\Omega)^{1/2}, \quad \eta_0 \equiv L/2z_0.$$

It suffices to retain in this equation one of the terms (its choice is determined by the sign of the transverse momentum of the particle on entering the field). Let, for the sake of argument, $p_\perp > 0$ and

$$|a_n|^2 = \frac{\eta_0/\pi}{(2n)^{1/2}}\left(\frac{\sin\chi}{\chi}\right)^2, \quad \chi \equiv (\beta-(2n)^{1/2})\eta_0. \tag{3}$$

An analysis of (3) shows that the level populations have an extremely strongly pronounced maximum near the value $n \approx \beta^2/2 \approx \varepsilon_\perp/\Omega$, with a width $\delta n \sim (\pi/\eta_0)(\varepsilon_\perp/\Omega)^{1/2}$. Using for $|a_n|^2$ a representation with the aid of a $\delta$-function

$$|a_n|^2 \approx \frac{\beta}{(2n)^{1/2}}\delta(n-\beta^2/2)$$

and changing from summation over n to integration, we find that the expression in the square brackets of (1) is equal to

$$[\,] = -f(\varepsilon_0) + \omega_1 \frac{df}{d\varepsilon} \frac{\varepsilon_\perp}{\Omega}. \tag{4}$$

The ratio of the absolute value of the first term to the second in (4) is defined by the parameter

$$\varsigma = \left| f(\varepsilon_0) / \frac{\omega_1 \varepsilon_\perp}{\Omega} \frac{df}{d\varepsilon} \right| \sim \frac{\delta\varepsilon \, \Omega}{\omega_1 \varepsilon_\perp}. \tag{5}$$

At $\varsigma > 1$, in accordance with the sign of the first term, the particles are accelerated on account of a predominant absorption of the wave quanta. On the contrary, at $\varsigma < 1$ the wave is amplified and the gain is determined in this case, as in the case of a magnetic undulator, by the sign and magnitude of the derivative $\frac{df}{d\varepsilon}$ (the recoil effect). The amplification condition imposes a lower limit on the transverse energy of the particles:

$$\varepsilon_\perp > \varepsilon_{\perp amp} = \frac{\delta\varepsilon \, \Omega}{\omega_1} \sim \frac{m_e}{\gamma} \frac{\delta\varepsilon}{\varepsilon}.$$

Assuming that the condition (4) is satisfied, we obtain from (1)

$$\Delta w^{(s=1)} = \frac{\pi}{4} \left( \frac{eA_{02}}{z_0} \right)^2 |\mathbf{e}_2 \mathbf{e}_z|^2 \frac{\gamma^2 \varepsilon_\perp}{\varepsilon \Omega} \frac{df}{d\varepsilon}, \quad (\varepsilon_\perp / \varepsilon)^{1/2} \gamma^2 \theta < 1. \tag{6}$$

We turn now to the limiting case $(n\alpha)^{1/2} \sim (\varepsilon_\perp / \varepsilon)^{1/2} s\gamma^2 \theta \gg 1$, when the dipole approximation cannot be used. Bearing in mind the known asymptotic representation of the Bessel function [60], we find that the number s should satisfy the condition $|2(n\alpha)^{1/2} - s| \leq s^{1/2}$, and Eq. (22) of [3] takes then the form

$$L_n^s(\alpha) \approx 0.53 n^s e^{\alpha/2} (n\alpha)^{-s/2 - 1/6} \quad (n\alpha)^{1/2} \gg 1, \quad n \gg s.$$

again, assuming the amplification condition (5) to be satisfied (the parameter is in the general case independent of s), we obtain from (21) of [3] the following expression for the difference between the total emission and absorption probabilities per unit time:

$$\Delta w^{(s)} = \frac{\pi}{2}(eA_{02})\frac{\gamma^2(df/d\varepsilon)}{\varepsilon(s/2)^{2/3}}\sum_n |a_n|^2 (0.53)^2 \left\{|\mathbf{e}_2\mathbf{e}_p| + (2n)^{1/2}\frac{|\mathbf{e}_2\mathbf{e}_z|}{z_0}\right\}^2. \tag{7}$$

For a wave with arbitrary polarization, as in the dipole-approximation case, the first term of the expression in the curly brackets of (7) is small if the condition $\theta < (\varepsilon_\perp/\varepsilon)^{1/2} \leq \Theta_0$ is satisfied. Omitting this term, we obtain from (7)

$$\Delta w^{(s)} \approx \frac{\pi}{4}\left(\frac{eA_{02}}{z_0}\right)^2 |\mathbf{e}_2\mathbf{e}_z|^2 \frac{\gamma^2(df/d\varepsilon)}{\varepsilon(s/2)^{2/3}}\frac{df}{d\varepsilon}\sum_n n|a_n|^2, \tag{8}$$

$$\left(\frac{\varepsilon_\perp}{\varepsilon}\right)^{1/2}\gamma^2\theta \geq 1.$$

The differences Aw of the total probabilities of emitting and absorbing per unit time a quantum of frequency o determine the gain Gat this frequency:

$$G = \frac{\Delta w \omega N_e}{E_2^2/8\pi} \tag{9}$$

where $N_e$ is the electron density in the beam and $E_2$ is the amplitude of the electric-field intensity in the amplified wave.

## 3. THE GAINS

Substituting Eqs. (6) and (8) in the definition (9) of G, and using the approximate equality (2), we obtain for the corresponding gains per pass the following expressions:

$$G^{(s=1)} = \frac{2\pi^2 e^2 |\mathbf{e}_2\mathbf{e}_z|^2 N_e l \varepsilon_\perp}{m_e^2 \omega_1}\left(\frac{\varepsilon}{\delta\varepsilon}\right)^2, \quad (\varepsilon_\perp/\varepsilon)^{1/2}\gamma^2\theta < 1; \tag{10}$$

$$G^{(s)} = \frac{2\pi^2 e^2 |\mathbf{e}_2\mathbf{e}_z|^2 N_e l \varepsilon_\perp}{m_e^2 \omega_s (s/2)^{1/3}}\left(\frac{\varepsilon}{\delta\varepsilon}\right)^2, \quad (\varepsilon_\perp/\varepsilon)^{1/2}\gamma^2\theta \geq 1 \tag{11}$$

(the numbers in (11) is determined by the value of the frequency $\omega$ of the amplified wave). We emphasize once more that Eqs. (10) and (11) are valid under the assumption that the transverse energy of the particle as it enters the field does not exceed the height of the potential barrier:

$\varepsilon_\perp \leq V_{0\perp}$. We note that the presented expressions (10) and (11) are universal and are equally applicable to an electrostatic and to a magnetostatic field (the contribution of the term $\sim (\mathbf{A}_1 \cdot \mathbf{A}_2)$ which enters in the expression for the operator of the perturbation for the constant magnetic field is small in the ratio

$$\Omega |\mathbf{e}_1 \mathbf{e}_2| / (\varepsilon \varepsilon_\perp)^{1/2} |\mathbf{e}_2 \mathbf{e}_z|.$$

The expressions (10) and (11) for the gains are valid in the limit $\delta\varepsilon/\varepsilon > 2\pi/\Omega$. A more realistic situation is one in which the inverse condition holds: $\delta\varepsilon/\varepsilon < 2\pi/\Omega$. In this case the spontaneous-emission line width is determined not by the energy scatter of the particles, but by the diffraction width connected with the fact that the region of interaction of the electrons with the field is finite. From the formal point of view the derivative $df/d\varepsilon \sim 1/(\delta\varepsilon)^2$ (36) and (11) should be replaced by

$$s^2 \frac{(l\Omega)^2}{\pi(\delta\varepsilon)^2} \left(\frac{\delta\varepsilon}{\varepsilon}\right)^2 \frac{d}{du} \frac{\sin^2 u}{u^2}, \quad s \leq s_0 = \frac{\varepsilon}{\delta\varepsilon} \frac{\pi}{l\Omega},$$

where $u = l\Omega(\varepsilon - \varepsilon_0)/\varepsilon_0$; the maxima of the gains correspond to $u = -1.5$. The gains should be calculated from the formulas

$$G^{(s=1)} = \frac{\pi^2 e^2 |\mathbf{e}_2 \mathbf{e}_z|^2 N_e l^3 \varepsilon_\perp}{\varepsilon^2 d} \left(\frac{2V_{0\perp}}{\varepsilon}\right)^{1/2} \frac{d}{du} \frac{\sin^2 u}{u^2}, \tag{12}$$

$$G^{(s=1)} = \frac{\pi e^2 |\mathbf{e}_2 \mathbf{e}_z|^2 N_e l^3 \varepsilon_\perp}{\varepsilon^2 d} \left(\frac{2V_{0\perp}}{\varepsilon}\right)^{1/2} \frac{d}{du} \frac{\sin^2 u}{u^2} s^{1/2}, \quad s \leq s_0. \tag{13}$$

We note that formula (12) coincides with the expression obtained in Ref. [61] for the electrostatic field by another method.

## 4. DISCUSSION OF RESULTS

We turn to an analysis of basic expressions obtained in the paper. As follows from (19) of [3], amplification of the wave at the frequency $\omega_1 = 2\gamma^2 \Omega$ is possible in the collinear scheme. When a wave is launched at an angle $\theta$ to the direction $\mathbf{p}_\parallel$ if $(\varepsilon_\perp/\varepsilon)^{1/2} \gamma^2 \theta \geq 1$ amplification is possible at the frequency $\omega_1 = 2\gamma^2 s \Omega$ ($\gamma^2 \theta^2 \ll 1$). Amplification at higher frequencies (on account of the increase of s) is accompanied, however, other conditions being equal, by a decrease of the gain: $G^{(s)} \sim G^{(s=1)} / s(s/2)^{2/3}$ $(s > s_0)$.

From the point of view of the effectiveness of the scheme considered in the paper, it is of interest to compare the gains obtained [Eqs. (12) and (13)] with the gains in a magnetic undulator. In this case, of course the gains must be compared at the same frequency. From the condition that the frequencies be equal we obtain the following relation between the parameters of the fields in the considered scheme and of the undulator (at equal electron energies):

$$\left(\frac{2K}{\gamma}\right)^{1/2} = \frac{2\pi d}{\lambda_u (1+K_u^2)}, \quad \left(\frac{K}{\gamma}\right)^{1/2} \ll 1, \tag{14}$$

where $K_u = eA_u / m_e$ is a parameter that characterizes the intensity of the electron interaction with the undulator field ($A_u$ is the amplitude of the vector potential of this field); $\lambda_u$ is the undulator period.

We use an equation given in Ref. [62] for the gain in a magnetic undulator in the collinear scheme:

$$G_0 = \frac{4\pi^2 e^2 N_e L_u^3 K_u^2}{\gamma m_e \omega (1+K_u^2) \lambda_u^2} \frac{d}{du} \frac{\sin^2 u}{u^2}. \tag{15}$$

for convenience in the comparison we rewrite the expression for $G^{(s=1)}$ [Eq. (12)] in comparable terms:

$$G^{(s=1)} = \frac{4\pi e^2 N_e l^3 \varepsilon_\perp K}{d^2 \gamma m_e^2 \omega} \frac{d}{du} \frac{\sin^2 u}{u^2}. \tag{16}$$

The ratio of the gains, as follows from (15) and (16) is equal to (we assume the regions. interaction, $l_u = l$)

$$G^{(s=1)} / G_u = 2\gamma \varepsilon_\perp / m_e K^2 (1+K_u^2). \tag{17}$$

It follows hence that in the case $K \sim K_u \sim 1$ and $\varepsilon_\perp \sim m_e$ gain in the scheme considered by us exceeds that of a magnetic undulator by a factor of the order of $\gamma$. The result can be understood by comparing the matrix:lements of the pertur- bation operators for the undulator, $\widehat{V}_u \sim e^2 (\mathbf{A}_1 \cdot \mathbf{A}_1)$ and for a field with transverse gradient $\widehat{V}_\perp \sim e^2 (\mathbf{A}_{02})_z (\partial/\partial z)$ in the collinear scheme. The ratio of the squares of these matrix ele which determine the gains, is of the order of

$$(V_u / V_\perp)^2 \sim \left( \frac{e^2 A_u A_{02} |\mathbf{e}_1 \mathbf{e}_2|}{(eA_{02}/z)|\mathbf{e}_1 \mathbf{e}_z| n^{1/3}} \right)^2 \sim \frac{(eA_u)^2}{\varepsilon_\perp \varepsilon} e^2.$$

If $\varepsilon_\perp \sim V_u = eA_u \sim m_e$, this ratio turns out to be of the order of $\gamma^{-1}$.

In conclusion, we present by way of example numerical estimates of the amplified-wave frequencies and of the gain G (s = 1) for a storage ring with the following beam and field characteristics: $\varepsilon$ = 150 MeV ($\gamma$ = 300); $\delta\varepsilon/\varepsilon = 0.5 \times 10^{-3}$; $N_e = 2 \times 10^{11} cm^{-3}$; d = 0 .2 cm; l = 1m (typical length of the linear section in a storage ring); $V_{0\perp} = 1 \times 10^6$ eV; $V_{0\perp}/d^2 \approx 60$ kG/cm; $\varepsilon_\perp = V_{0\perp}/2$. At the parameters chosen by us the inequality $\delta\varepsilon/\varepsilon < 2\pi/\Omega$ is satisfied, therefore the gain is calculated using Eq. (12). The value of the amplified frequency for these parameters is $\omega_1 \approx 1.8$ eV, and the gain at this frequency is $G^{(s=1)} \approx 0.7$.

The indicated field characteristics can be realized by a set of quadrupole (multipole) magnetic lenses made, e.g., of magnets based on a samarium-cobalt alloy. In the case of an electrostatic field the field configuration considered in the paper can be obtained between a system of parallel plates that carry in vacuum charges of like sign (negative for electrons). The use of magnetic lenses, however, is preferable since the electrostatic field intensities are limited by the autoionization effect.

Stimulated emission of ultrarelativistic electrons in electric (magnetic) fields with large transverse field gradients makes it thus possible, under certain conditions, to obtain noticeably larger gains than in ordinary magnetic undulators. In this case we are dealing with optical and near-ultraviolet frequencies. The possibility of generating harder coherent radiation calls for a separate analysis.

In the derivation of the equations we used a quantum treatment of the electron motion in the field. However, the gains (12) and (13) do not depend on the Planck constant $\hbar$ and can therefore be obtained within the framework of a classical description (see [42]).